\documentclass[journal,onecolumn,10pt,a4paper]{IEEEtran}
\usepackage[english]{babel}
\usepackage[utf8]{inputenc}
\usepackage{amsfonts,amsmath,amsthm,amssymb}
\usepackage{cancel}
\usepackage{graphicx}
\usepackage{float}
\usepackage{subfigure}

 \title{Adaptive Transmission Techniques for Mobile Satellite Links}

 \author{
  Jesús Arnau\thanks{Email: suso@gts.uvigo.es \ \ \ \ \ \ \ \ Phone: +34 986 818737  Fax: +34 986 812116.}
\ ,
  Alberto Rico-Alvariño\thanks{Email: alberto@gts.uvigo.es \ \ \  \ Phone: +34 986 818737 Fax: +34 986 812116.}
  \ and Carlos Mosquera\thanks{Email: mosquera@gts.uvigo.es  \ Phone: +34 986 812677  Fax: +34 986 812116.}\\
  {\normalsize\itshape
   Signal Theory and Communications Department, University of Vigo - 36310 Vigo, Spain.} \\ 
}

\newcommand{\field}[1]{\mathbb{#1}}

\def\0{{\mathbf 0}}
\def\1{{\mathbf 1}}

\def\h{{\mathbf h}}

\def\x{{\mathbf x}}

\def\gammaeff{{\gamma_\textnormal{eff} }}

\def\gammab{{\boldsymbol \gamma}}

\newcommand{\ex}{\mathbb{E}}

\newcommand{\openone}{\leavevmode\hbox{\small1\normalsize\kern-.33em1}}

\begin{document}

\maketitle

\begin{abstract}
Adapting the transmission rate in an LMS channel is a challenging task because of the relatively fast time variations, of the long delays involved, and of the difficulty in mapping the parameters of a time-varying channel into communication performance. In this paper, we propose two strategies for dealing with these impairments, namely, multi-layer coding (MLC) in the forward link, and open-loop adaptation in the return link. Both strategies rely on physical-layer abstraction tools for predicting the link performance. We will show that, in both cases, it is possible to increase the average spectral efficiency while at the same time keeping the outage probability under a given threshold. To do so, the forward link strategy will rely on introducing some latency in the data stream by using retransmissions. The return link, on the other hand, will rely on a statistical characterization of a physical-layer abstraction measure.
\end{abstract}
\section*{Nomenclature}
\begin{tabbing}
  XXXXXXX \= \kill
  ACM \> Adaptive Coding and Modulation\\
  ARQ \> Automatic Repeat Request\\
  CDF \> Cumulative Distribution Function\\
  CLT \> Central Limit Theorem\\
  CSI \> Channel State Information\\
  ESM \> Effective SNR Mapping\\
  GEO \> Geostationary\\
  ITS \> Intermediate Tree Shadowed\\
  LMS \> Land Mobile Satellite\\
  LOS \> Line-of-Sight\\
  MIESM \> Mutual Information ESM\\
  MODCOD \> Modulation and Coding\\
  MLC \> Multi-Layer Coding\\
  PDF \> Probability Density Function\\
  SNR \> Signal-to-Noise Ratio\\
  WER \> Word Error Rate\\
  
 \end{tabbing}

\section{Introduction}\label{sec:intro}
In mobile satellite communications, there is an increasing need for more efficient transmission techniques that enable higher bit-rates at an affordable cost. To this extent, Adaptive Coding and Modulation (ACM) allows the provision of broadband services to large user populations at lower costs, since it makes it possible to operate the links more efficiently by selecting the most suitable Modulation and Coding Scheme (MODCOD) at each time \cite{bischl10}. However, the use of ACM for mobile links operating at S-band is hindered by the behavior of the Land Mobile Satellite Channel (LMS)\cite{fontan01}. This channel is usually modeled by a fast fading component, whose spectrum is related to the mobile speed by the Doppler effect, superimposed on a slow shadowing component; the parameters of both fading and shadowing depend on the environment in which the receiver happens to be. In short, the mobility of the user terminal will cause fast, difficult to predict channel variations, which will pose additional difficulties on the design of both forward and return link strategies.

Elaborating more on this issue, adaptation can be performed in {\it open-loop} or {\it closed loop}. In an open-loop scheme, the transmitter directly measures the signal quality from the other link and changes the parameters accordingly; it usually enjoys negligible delays, although at the price of having only partial information when both links are not perfectly correlated. On the other hand, closed-loop strategies wait for the other end to process their data and operate upon receiving some information about its reception. This makes them more accurate, although the experienced delay is much higher.

Another key problem is that it is difficult to relate the channel statistics --even if we knew them-- to the performance of the link because the channel is time varying. For example, it is difficult to tell what is best, a fixed channel with low quality or a fast varying channel with sharp transitions and a higher average quality. In an LMS channel, where transmission usually entails the use of complicated channel codes, if is difficult to do this without resorting to extensive simulations. This, however, is highly undesired for the design of an ACM strategy. In order to overcome this problem, we propose to use physical layer abstraction techniques. Thus, we will use Effective SNR Metrics (ESM) that have been reported to map complicated channel profiles with their actual performance in a one-to-one manner\cite{wimax_rbir}. Therefore, our key assumption will be the following: {\it a codeword will be correctly transmitted if the ESM of the channel that it undergoes is higher than or equal to the decoding threshold specified for the MODCOD in use in a static channel}.

Focusing firstly on the forward link, the fading is attached to the receiver and, as a consequence, the delay experienced by the CSI will be much longer than the channel coherence time for most speeds of practical interest. Therefore, the received CSI will be completely outdated --even if considering an open-loop scheme\cite{monk1995}-- and adaptive rate will be of no use. However, if frames can be retransmitted or, at least, some additional redundancy or parity bits can be sent, then higher throughputs can be achieved at the cost of some latency. The underlying idea would be to transmit at a higher rate during good states of the channel, while somehow keeping the outage probability low. To achieve this, we propose to resort to superposition or multilayer coding (MLC) with retransmissions, in a spirit close to Hybrid-ARQ systems or Layered Rateless Codes\cite{Wornell08}. Particularly, we consider the use of two layers of information, each of them modulated at a different power. Upon reception, a decoding performance indicator will be sent and those bits which were not decoded by the receiver will be resent in a robust way (that is, in the high power layer) in order to keep the latency low. 

On the other hand, the return link enjoys timely channel information when operating in open-loop mode, but at the cost of having only partial CSI. This is so because, if both links operate on different frequency bands, then only the shadowing component can be accurately estimated; the fast fading will be uncorrelated and only some of its statistical properties will be inferred. To cope with this, we propose an ACM algorithm in which the transmitter selects the most efficient MODCOD that guarantees a given outage probability based upon the prediction of the link performance. This prediction is based on approximating the ESM by a log-normally distributed random variable whose parameters can be deduced from open-loop estimation. 

In this paper, we will describe the mathematics and assumptions supporting both algorithms, along with a detailed description of both of them. In what refers to the forward link, we will show how to design such a multilayer scheme, based upon stale CSI, in a way that ensures a target outage probability while exploiting the good states of the channel to transmit at a higher rate. Simulation results will show the performance achieved by the proposed schemes against those provided by more conventional, less adaptive techniques. Special attention will be paid to the availability of the link during the most compelling states of the channel, illustrating the achievable performance and the trade-offs involved. 

The remaining of the document is structured as follows: Section~\ref{sec:sys_model} describes the system under study, focusing on the mathematical channel model and on a brief introduction to physical layer abstraction; Section~\ref{sec:fl} is devoted to the design and simulation of an MLC scheme for the forward link; Section~\ref{sec:rl} describes the solution and the results obtained for the return link; finally, conclusions are summarized on Section~\ref{sec:conclusions}.

\section{System model}\label{sec:sys_model}
\begin{figure}
 \centering
 \includegraphics[width=.9\columnwidth]{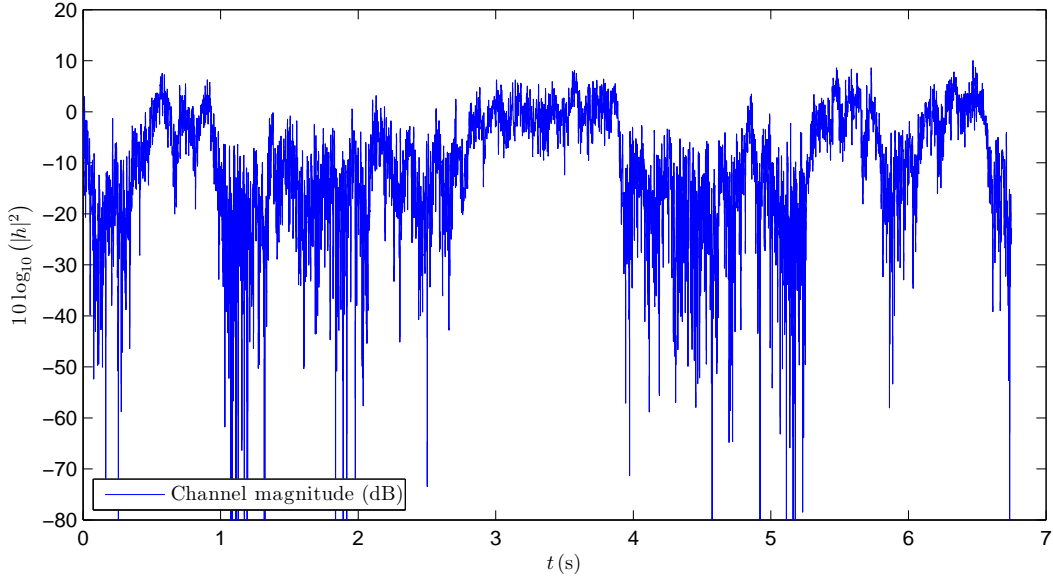}
 \caption{Evolution of the channel SNR with respect to time, $f = 2$\,GHz, $v = 30$\,m/s, intermediate tree shadowed (ITS) area.}
 \label{fig:channel}
\end{figure}

We will assume a mobile geostationary (GEO) satellite channel, operating at the S-band, in which transmissions will be structured in codewords of $L$ symbols each, with $T_{\textnormal{symb}}$ the symbol period. As a consequence, each burst will go through an LMS channel $\h = \left[ h_1 \ h_2 \ \cdots \ \ h_L\right]^T$ with
\begin{equation}
 \h = \h_{\textnormal{LOS}}+\h_{\textnormal{NLOS}}.
\end{equation}
Here, we have written separately the contribution of the log-normal component $\h_{\textnormal{LOS}}$ and the Rayleigh contribution $\h_{\textnormal{NLOS}}$. Let $\sigma^2$ be the noise power and  $\gammab = \left[ \gamma_1 \ \gamma_2 \ \cdots \ \ \gamma_L\right]^T$ the SNR experienced by each set of $L$ symbols after going through the LMS channel, with $\gamma_i = |h_i|^2/\sigma^2$. 

One of the most common models for the LMS channel is the so-called 3 state Fontan channel model \cite{fontan01}. In this model,
line-of sight (LOS), small, and heavy shadowing conditions are taken into account by three different states characterized  by a first-order Markov chain whose state and transition probabilities were extracted from experiments and test
campaigns (see Fig.~\ref{fig:channel} for an example of the channels behavior). Each of the states has a correlation length of $3$--$5$\,m in the S-band, and its variations
depend primarily on the relative speed between transmitter and receiver. Focusing on a specific state, the fading behavior follows the Loo model \cite{loo1985}: slow variations are described by a log-normal distribution, whereas fast fluctuations of the signal amplitude are given by a Rician distribution. Mathematically speaking, the probability density function (pdf) of the signal amplitude at a given time instant would be given by

\begin{equation}
 f_r(x) = \frac{x}{b_0\sqrt{2\pi d_0}}\int_0^\infty\frac{1}{z}\exp\left(-\frac{(\log z-\mu)^2}{2d_0}-\frac{x^2+z^2}{2b_0}\right) I_0\left(\frac{x\cdot z}{b_0}\right)\ \partial z
\end{equation}
where $d_0$ and $\mu$ are the scale parameter and the location parameter of the log-normal distribution, respectively.

In the case of the complete Fontan channel model, the pdf of the signal amplitude has to be weighted by the probabilities of the different states, given a certain environment, thus resulting in a \textit{mixture} of Loo pdfs
\begin{equation}
\label{e:looMixture}
 f_{r,F}(x) = \sum_{i=1}^N p_i \frac{x}{b_0\sqrt{2\pi d_{0,i}}}\int_0^\infty\frac{1}{z}\exp\left(-\frac{(\log z-\mu_i)^2}{2d_{0,i}}-\frac{x^2+z^2}{2b_{0,i}}\right) I_0\left(\frac{x\cdot z}{b_{0,i}}\right)\ \partial z
\end{equation}
with $i$ an index that is spanning the $N$ different states of the Markov chain, $p_i$ the probability of the $i$-th state, and $d_{0,i}$, $\mu_i$ and $b_{0,i}$ the parameters of the Loo distribution in the $i$-th state.

It is important to remark that the different $h_i$ will not be independent. In fact, $\h_{\textnormal{NLOS}}$ can be seen as the result of filtering complex white Gaussian noise with a low-pass filter with cut-off frequency given by the Doppler spread $f_D$. 

If the channel varies within the length of a codeword, then we need tools to tell whether each codeword will be correctly decoded or not. Therefore, our goal is quite similar to the one pursued by Physical Layer Abstraction techniques, that is, to obtain a metric -that we can compute from estimations of some parameters- which must be related in a one-to-one manner to the Word Error Rate (WER); such metrics are often called Effective SNR Mappings (ESM). The ESM metric in which we will focus is the Mutual Information Effective SNR Mapping (MIESM)\cite{link_performance}, which must be parametrized in terms of just the constellation used, not the code. It reads as
\begin{equation}
  \gamma_{eff} = \frac{1}{M}\left( \frac{1}{N}\sum_{n=1}^N\Phi\left(\gamma_n\right)\right)
\end{equation}
where $M$ is the number of points in the constellation and $\gamma_n$ is the SNR experienced by the $n$-th codeword. The metric $\Phi$ is usually approximated by a sum of exponentials. The simplest one would be

\begin{equation}\label{eq:ap1}
\Phi(x) \approx 1-e^{-\beta x}
\end{equation}
which can  be seen to be equivalent to the so-called Exponential ESM (EEESM)\footnote{Note that mapping one minus the mutual information is the same as mapping the mutual information as long as we select the appropriate inverse transformation.}. Moreover, the following approximation has been reported to show a remarkably good fit to the original function:
\begin{equation}\label{eq:ap2}
  \Phi(\gamma) \approx 1-\alpha_1e^{-\beta_1\gamma}+(1-\alpha_1)e^{-\beta_2\gamma}.
\end{equation}
In either case, parameters $\alpha$ and $\beta$ must be tuned for each constellation.
\section{Adaptive Rate in the Forward Link}
\label{sec:fl}

The main constraint of the forward link is that, even under open-loop operation, the total delay will be of $0.5$\,s. Since the fading event will be attached to the mobile terminal, the returning signal will experiment $0.25$\,s of delay since it undergoes the channel fading until it reaches the gateway; then, at least other $0.25$\,s will pass until the newly generated signal for the forward link reaches the mobile terminal. The effect of such large delays may be dramatic for the design of an ACM control strategy since, for many speeds of practical interest, the channel will be completely uncorrelated after those $0.5$\,s. As an example, Table \ref{tbl:sh_corr} shows the shadowing coherence time and the state 2 average length for different speeds and two environments: open and intermediate tree shadowed (ITS). As we can see, the channel could even have changed its Markov chain state after $0.5$\,s if the mobile terminal moves at a speed close to $15$\,m/s ($54$\,km/h).

Still, the good states of the channel can be used for transmitting at a higher rate, while guaranteeing a given outage probability. We propose to do so by resorting to multi-layer coding and splitting the information
to convey into two modes, one of them much more protected than the other; this would
be accomplished by the appropriate power balance between the two layers. The overall technique
would work by retransmitting in the most protected mode those codewords which had
failed before. 

In this section, we will present a theoretical framework for such multilayer transmission scheme. The optimal MODCOD and power balance will be found by numerical optimization for each environment in the Fontan model, and results will show a remarkable increase in spectral efficiency while keeping the outage probability low.
\begin{table}
  \caption{Shadowing correlation length (left) and state 2 average length (right), as a function of the mobile speed, for different environments.}
  \label{tbl:sh_corr}

  \centering
  \begin{tabular}{|l | c | c | c |} \hline
Env./Speed & 1 m/s & 15 m/s & 40 m/s \\ \hline
Open	& 2.5 s & 160 ms & 63 ms \\ \hline
ITS	& 1.5 s & 100 ms & 38 ms \\ \hline

  \end{tabular}\hspace*{1cm}
  \begin{tabular}{|l | c | c | c |} \hline
Env./Speed & 1 m/s & 15 m/s & 40 m/s \\ \hline
Open	& 7.5 s & 500 ms & 188 ms \\ \hline
ITS	& 6.3 s & 420 ms & 150 ms \\ \hline

  \end{tabular}
\end{table}

\newcommand{\doc}{report }
\newcommand{\Lalb}{{\cal L} }
\newcommand{\Halb}{{\cal H} }
\newcommand{\malb}[1]{\boldsymbol{\mathbf{#1}}}

\subsection{System Model for the Forward Link}

\begin{figure}
\begin{center}

 \includegraphics[width=.75\textwidth]{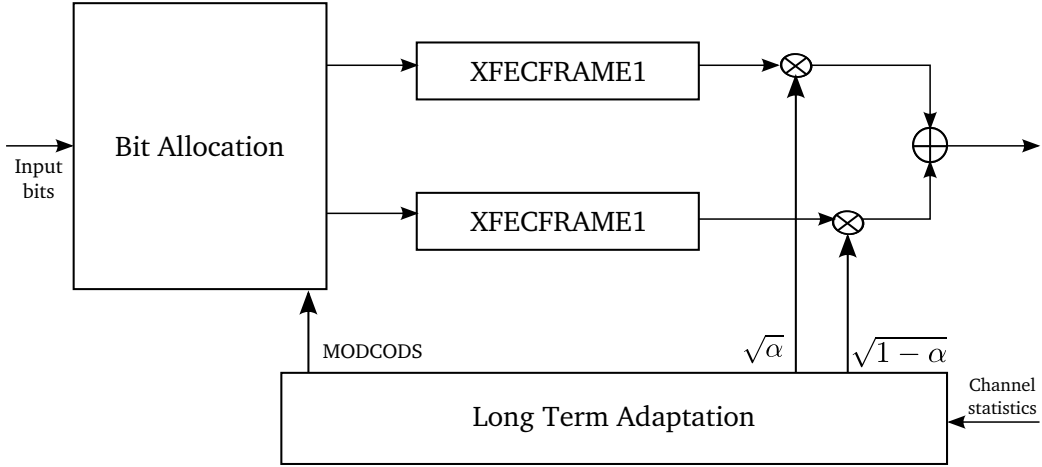}
\caption{Schematic of the MLC scheme: the input bits are divided into two different layers, and the transmission parameters are selected according to long-term channel statistics.}
\label{f:MLT}\end{center}
\end{figure}

If the transmitter uses Multi-Level Coding (MLC), the received signal can be written for the two-layer case as
\begin{equation}
 y_i = h_i \left( \sqrt{\alpha} x_i^{\cal H} + \sqrt{1-\alpha} x_i^{\cal L} \right) + n_i 
\end{equation}
with $n_i \sim {\cal CN}\left( 0, \sigma^2\right)$ a sample of white Gaussian noise, $h_i$  the (complex-valued) channel seen by the $i$-th symbol, $x_i^\Halb$ and $x_i^{\cal L}$ the $i$-th symbols of the ${\cal H}$ (High priority) and ${\cal L}$ (Low priority) levels, and $\alpha$ a parameter that weights the power sharing between the two layers. Note that for $\alpha=1$ we have a single-layer transmission. This MLC scheme is depicted in Figure \ref{f:MLT}.

Each codeword spans $L$ different channel states\footnote{States must not be confused with the LMS Markov model states.}, such that the $k$-th codeword is affected by the channel state vector $\malb h_k = \left[h_{kL+1}\,...\,h_{(k+1)L} \right]^T$. The channel state vector $\malb h_k$ is assumed to be distributed following a pdf $f_{\malb h}(\malb h_k)$ that cannot be decomposed (in general) into $f_{\malb h}(\malb h_k) = \prod_{i=1}^L f_{h} (h_{kL+i})$ due to the existence of correlation among the different states spanned by the same codeword.

For the sake of clarity, we define the Signal to Noise Ratio (SNR) of the $i$-th symbol of the $\Halb$ and $\Lalb$ levels as
\begin{equation}
\label{e:rhoH}
 \gamma_i^\Halb = \frac{\left| h_i \right| ^2 \alpha}{\sigma^2 + \left(1-\alpha\right) \left| h_i \right|^2}
\end{equation}
\begin{equation}
\label{e:rhoL}
 \gamma_i^\Lalb = \frac{\left| h_i \right| ^2(1-\alpha)}{\sigma^2}.
\end{equation} where the transmitted power is normalized to one.

We will consider that the $k$-th codeword corresponding to the $\Halb$ layer when using the $m$-th MODCOD is correctly decoded if the average Mutual Information (MI) between $\malb y_k = \left[ y_{kL+1}\,...,\, y_{(k+1)L}\right]^T$ and $\malb x_k^\Halb = \left[x_{kL+1}^\Halb\,...,\, x_{(k+1)L}^\Halb \right]^T$ is greater than or equal to a given rate threshold or, equivalently, that the effective SNR $\gamma_{eff}\left( k \right)$ of the $k$-th codeword is larger than the required SNR for the decodability of the selected $m$-th MODCOD $\gamma^{th}_m$:
\begin{equation}
  \Phi^{-1}\left(\frac{1}{L}\sum_{i={kL+1}}^{(k+1)L} \Phi^{{\cal M}(m)} \left(\gamma_i^\Halb \right)\right) \geq \gamma^{th}_m
\end{equation}
with $\Phi^{\cal M} ( \gamma)$ the mutual information between the input and output of a Gaussian channel with SNR $\gamma$ for an input restricted to the constellation ${\cal M} = \left\{m_1,\, ...\, m_{|{\cal M}|} \right\} \in {\field C}^{|{\cal M}|}$. Note that this approach is conservative, as the interference caused by the $\Lalb$ layer to the $\Halb$ layer is treated as Gaussian noise, when it is clear that it has a lower entropy, thus leading to a higher MI.

The $\Lalb$ layer will be correctly decoded if the $\Halb$ layer was correctly decoded and the MI between $\malb y_k - \malb x^\Halb_k$ and $\malb x^\Lalb_k = \left[x_{kL+1}^\Lalb\,...,\, x_{(k+1)\Lalb}^L \right]^T$ is greater than or equal to the rate threshold or, in terms of effective SNR, if the following condition is met:
\begin{equation}
 \left(\Phi^{-1}\left(\frac{1}{L}\sum_{i={kL+1}}^{(k+1)L} \Phi^{{\cal M}(k)} \left(\gamma_i^\Halb \right)\right) \geq \gamma_{m}^{th} \right) \cap  \left(\Phi^{-1}\left(\frac{1}{L}\sum_{i={kL+1}}^{(k+1)L} \Phi^{{\cal M}(k)} \left(\gamma_i^\Lalb \right)\right)  \geq \gamma_{m}^{th} \right).
\end{equation}

Note that we are restricting both layers to have the same constellation and code rates, which is desirable for implementation issues. In the following, we will restrict our analysis to the first codeword $k=0$, without loss of generality, and drop the indexes that take into account the codeword number.

\subsection{Problem Statement}

Our target is to maximize the average spectral efficiency subject to an outage probability constraint on the ${\cal H}$ layer. Let us define ${\cal C}_i = \left\{ {\cal M}_i, {\cal R}_i, {\gamma}_i^{th}\right\}$ as the triplet (modulation, spectral efficiency, required SNR for decodability) that defines the $i$-th MODCOD. We will restrict our modulation and code rates to the finite length set $\malb {\cal C} = \left\{ {\cal C}_k \right\}_{i=1}^{|\malb {\cal C}|}$. The parameters of the different MODCODs are summarized in Table \ref{tbl:modcod}. Note that the selected MODCODs for the forward link are those of DVB-S2 which use a QPSK constellation.

\begin{table}
  \caption{MODCOD parameters for the FL (left) and RL (right).}
  \label{tbl:modcod}

  \centering
  \begin{tabular}{|c | c | c |} \hline
MODCOD & ${\cal R}$ & ${\gamma}^{th}$ (dB) \\ \hline \hline
QPSK 1/4 & 0.357 & -1.5 \\ \hline
QPSK 1/3 & 0.616 & -0.3 \\ \hline
QPSK 2/5 & 0.745 & 0.6 \\ \hline
QPSK 1/2 & 0.831 & 1.9 \\ \hline
QPSK 3/5 & 1.132 & 3.1 \\ \hline
QPSK 2/3 & 1.261 & 4 \\ \hline
QPSK 3/4 & 1.390 & 4.9 \\ \hline
QPSK 4/5 & 1.476 & 5.6 \\ \hline
QPSK 5/6 & 1.562 & 6.1 \\ \hline
QPSK 8/9 & 1.691 & 7.1 \\ \hline
  \end{tabular}\hspace*{1cm} 
  \begin{tabular}{|c | c | c |} \hline
MODCOD & ${\cal R}$ & ${\gamma}^{th}$ (dB) \\ \hline \hline
QPSK 1/3 & 0.563 & 1.7\\ \hline
QPSK 1/2 & 0.874 & 4\\ \hline
QPSK 2/3 & 1.259 & 5.9\\ \hline
QPSK 3/4 & 1.422 & 7 \\ \hline
QPSK 5/6 & 1.600 & 8.3 \\ \hline
8PSK 2/3 & 1.704 & 9.9\\ \hline
8PSK 3/4 & 1.926 & 11.5\\ \hline
8PSK 5/6 & 2.197 & 13.1\\ \hline
16QAM 3/4 & 2.593 & 13.7 \\ \hline
16QAM 5/6 & 2.874 & 15.2\\ \hline
  \end{tabular}
\end{table}

Our objective function will be the Average Spectral Efficiency (ASE), defined as
\begin{equation}
\eta\left({\cal C}_k, \alpha \right) \dot= 
{\cal R}_k\cdot\left( P\left[ \gamma_{eff}^{{\cal H}} \geq \gamma_k^{th} \right] + P\left[ \gamma_{eff}^{{\cal H}} \geq \gamma_k^{th} \cap \gamma_{eff}^{{\cal L}} \geq \gamma_k^{th}\right] \right)
\end{equation}
with $P\left[ A \right]$ the probability of occurrence of event $A$, and
\begin{equation}
 \gamma_{eff}^{{\cal T}} \dot = \Phi^{-1}\left(\frac{1}{L}\sum_{i=1}	^{L} \Phi^{{\cal M}(k)}\left( \gamma_i^{{\cal T}}\right)\right)
\end{equation}
for ${\cal T} \in \left\{ {\cal L}, {\cal H}\right\}$. The outage probability constraint is defined as
\begin{equation}
 g\left({\cal C}_k, \alpha \right) \dot = P\left[ \gamma_{eff}^{\cal H} < \gamma_k^{th} \right] \leq P_{out}.
\end{equation}
Therefore, our design problem is stated as
\begin{equation}
\label{e:optProb}
  \begin{array}{ll}
  \mbox{maximize } & \eta\left({\cal C}_k, \alpha \right) \\
  \mbox{subject to } & g\left({\cal C}_k, \alpha \right) \leq P_{out} \\
		     & 0 \leq \alpha \leq 1
\end{array}		
\end{equation}
where the maximization is performed over ${\cal C}_k \in \malb {\cal C}$ and $\alpha \in {\field R}$. The MODCOD choice is assuming the knowledge of the channel statistics, which we consider to be available at the gateway.

In the following, we will analyze the problem  by assuming a block fading channel, which is a realistic assumption if the mobile speed is small enough, and makes the problem analytically tractable, and afterwards evaluate the evolution of throughput and outage probability as a function of the mobile speed.

\subsection{Block Fading}
\begin{figure}
\begin{center}
 \includegraphics[width=.6\textwidth]{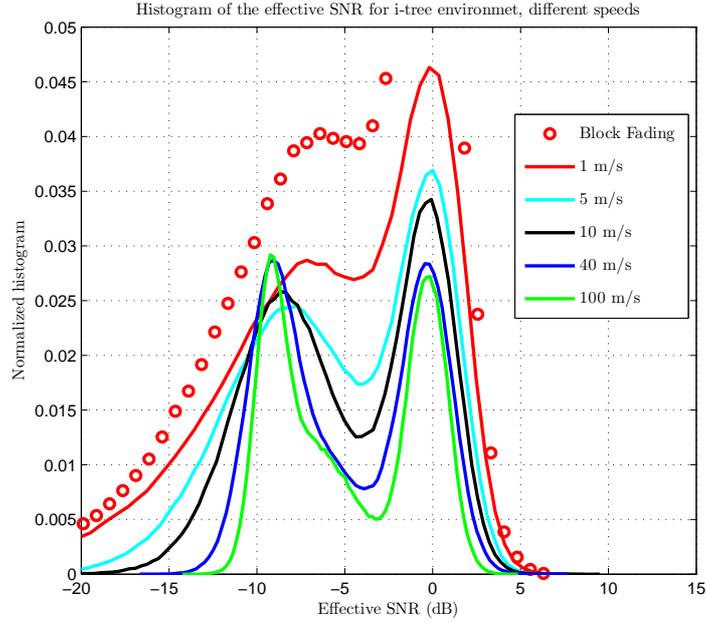}
\caption{Histogram of the effective SNR for different speed values, ITS environment, very low SNR. }
\label{s1}
\end{center}\end{figure}

The histogram of the effective SNR in a Fontan channel has been obtained for different mobile speeds and for different \textit{average SNR} values. 
In Figure \ref{s1} we can see the variation of the effective SNR with speed. Clearly, as the speed increases the effective SNR variance diminishes as the result of averaging more \textit{channel states} in the same codeword.

Unfortunately, obtaining the pdf of the effective SNR for a mobile seems to be analytically intractable, so we will assume that all the channel states seen by a codeword are approximately the same ({\it block fading}): 
\begin{equation}
 h_1 \approx h_k,\, k=2,\,...\, L
\end{equation}
so we can approximate
\begin{equation}
 \gamma_{eff}^{\cal H} \approx \gamma^{\cal H}
\end{equation}
with $\gamma^{\cal H} \approx \gamma_1^{\cal H},\,...,\,\gamma_L^{\cal H}$.
If this is the case we can define the event of decodability of the $\Halb$ layer as
\begin{equation}
\label{e:decH}
 \gamma^\Halb \geq \gamma_k^{th}.
\end{equation}
Analogously, we can rewrite the event of decodability of the $\Lalb$ layer as
\begin{equation}
\label{e:decL}
 \left( \gamma^\Halb \geq \gamma_k^{th }\right) \cap \left( \gamma^\Lalb \geq \gamma_k^{th }\right).
\end{equation}
Following (\ref{e:rhoH}) we can rewrite (\ref{e:decH}) as
\begin{equation}
 \left| h \right|^2 \geq \frac{\sigma^2 \gamma_k^{th}}{\alpha - \left( 1-\alpha\right)\gamma_k^{th}},
\end{equation}
and (\ref{e:decL}) according to (\ref{e:rhoL})  as
\begin{equation}
 \left| h \right|^2 \geq \max \left\{\frac{\sigma^2 \gamma_k^{th}}{\alpha - \left( 1-\alpha\right)\gamma_k^{th}}, \frac{\gamma_k^{th}\sigma^2}{1-\alpha} \right\}.
\end{equation}
Note that if we choose a value of $\alpha$ such that 
\begin{equation}
 \frac{\sigma^2 \gamma_k^{th}}{\alpha - \left( 1-\alpha\right)\gamma_k^{th}} > \frac{\gamma_k^{th}\sigma^2}{1-\alpha}
\end{equation}
then the ${\cal L}$ layer will be only limited by the decodability of the $\Halb$ layer, which is clearly not optimum. Therefore, we might want to choose a value of $\alpha$ such that
\begin{equation}
 \alpha \geq \frac{\gamma_{k}^{th}+1}{\gamma_{k}^{th}+2}.
\end{equation}
By adding this constraint to $\alpha$, the ASE can be written as
\begin{equation}
 \eta\left({\cal C}_k, \alpha \right) = \mathcal{R}_k\cdot \left( P\left[  \left| h \right|^2 \geq \frac{\sigma^2 \gamma_k^{th}}{\alpha - \left( 1-\alpha\right)\gamma_k^{th}} \right] + P\left[ \left| h \right|^2 \geq \frac{\gamma_k^{th}\sigma^2}{1-\alpha} \right] \right)
\end{equation}
and the outage probability
\begin{equation}
\label{e:constBl}
 g\left({\cal C}_k, \alpha \right) = P\left[  \left| h \right|^2 \leq \frac{\sigma^2 \gamma_k^{th}}{\alpha - \left( 1-\alpha\right)\gamma_k^{th}} \right] \leq P_{out}.
\end{equation}
Note that the outage probability applies to the ${\cal H}$ layer and, in consequence, $g\left({\cal C}_k, \alpha \right)$ is a monotonic decreasing function of $\alpha$, as it is clear that allocating more power to the $\Halb$ layer will decrease the outage probability. Therefore, if we assume that $g\left({\cal C}_k, 1 \right) \leq P_{out}$ (otherwise the problem will be infeasible for the MODCOD ${\cal C}_k$), constraint (\ref{e:constBl}) is equivalent to
\begin{equation}
 \alpha \geq \alpha^{out}_k
\end{equation}
with $\alpha^{out}_k$ such that $g\left({\cal C}_k, \alpha^{out}_k \right) = P_{out}$. Therefore, for a given ${\cal C}_k$ such that the problem is feasible, the optimum value $\eta^\star_k$ is obtained as
\begin{equation}
 \eta^\star_k = \max_{\alpha^{min}_k \leq \alpha \leq 1} \left\{  \eta\left({\cal C}_k, \alpha \right) \right\}
\end{equation}
with
\begin{equation}
 \alpha^{min}_k \dot = \max \left\{\alpha^{out}_k,\, \frac{\gamma_{k}^{th}+1}{\gamma_{k}^{th}+2}  \right\},
\end{equation}
\begin{equation}
 \eta^\star = \max\left\{ \eta^\star_k \right\}_{k=1}^{|\malb{\cal C}|}.
\end{equation}
Note that the previous expressions involving probabilities of the channel power $\left| h \right|^2$ being smaller than a given threshold $k$ can be easily rewritten as
\begin{equation}
 P\left[  \left| h \right|^2 \leq k \right] = F_{r,F}\left( \sqrt{k}\right)
\end{equation}
where
\begin{equation}
 F_{r,F}\left( \sqrt{k}\right) = \int_{x=0}^{\sqrt{k}} f_{r,F}\left( x \right) \partial x,
\end{equation}
being $f_{r,l}$ the pdf of the Loo mixture (\ref{e:looMixture}).


%

\begin{figure}
\begin{center}

 \includegraphics[width=.6\textwidth]{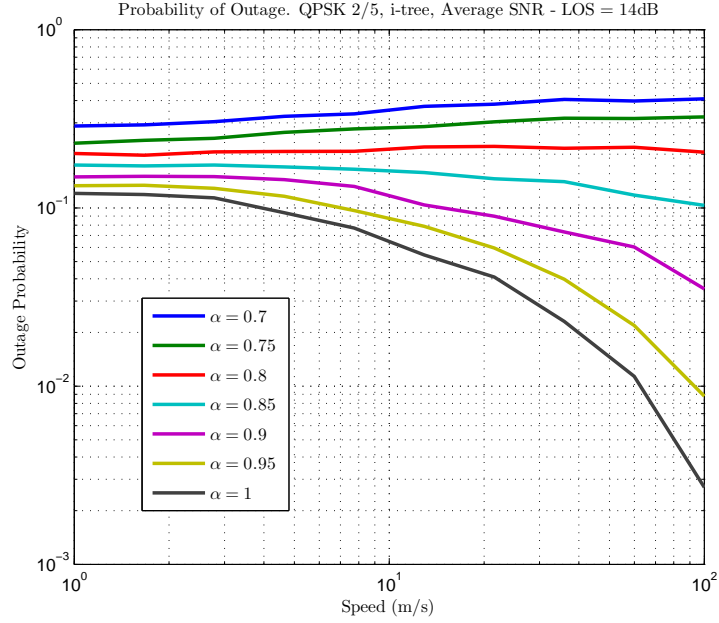}
\caption{Outage probability for different values of $\alpha$ and speed.}

\label{outProbabilityMultAlpha}\end{center}
\end{figure}

%
%
%
%

\subsection{Evolution of Outage Probability with Speed and Environment}
It is clear that for a given Line of Sight SNR (the SNR a receiver would experience without shadowing and fading), the performance of a receiver (in terms of ASE or outage probability) is going to be heavily dependent on the environment, which determines the fading and shadowing statistical characterization. Moreover, as we have already seen, an increasing mobile speed causes different channel states to be \textit{averaged} during the same codeword, thus reducing the variance of the effective SNR and, as a consequence, reducing the outage probability.

Therefore, as we are constraining the outage probability to lie below a given threshold (according to (\ref{e:optProb})), it would be useful to analyze the outage probability performance of the most protected MODCOD for different speeds and environments. Note that if the most protected MODCOD does not provide the desired outage probability, then it is clear that the optimization problem (\ref{e:optProb}) is unfeasible.

In order to anticipate the performance of the MLC strategy for the LMS channel, we have simulated its performance for different values of $\alpha$ and speed, as we can see in Figure \ref{outProbabilityMultAlpha}. Obviously, the larger $\alpha$ is, the less interference we receive from the ${\cal L}$ layer, so the outage probability decreases as $\alpha$ increases. Moreover, we have seen in Figure~\ref{s1} that the effect of incrementing the speed is to reduce the variance of the ESM, as more states are averaged for the same codeword. As an outage event occurs when the ESM of a codeword is very small, the variance reduction implies that this event is less likely. As a consequence, the outage probability decreases with speed.

\begin{figure}
\hspace{-.7cm}
\subfigure[]{\includegraphics[width=.57\textwidth]{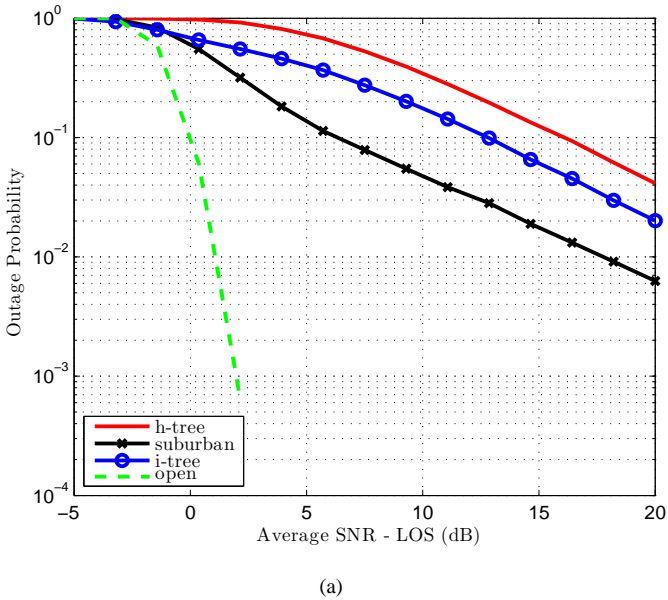}}\hspace*{-0.9cm}
\subfigure[]{\includegraphics[width=.57\textwidth]{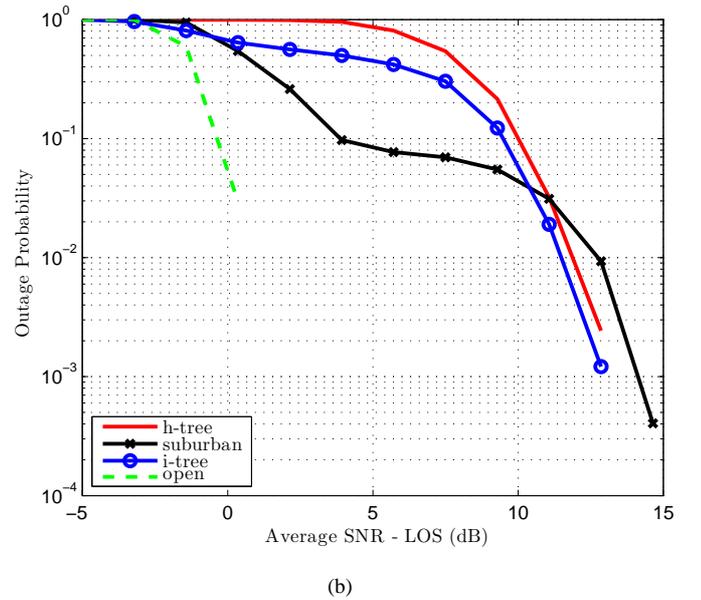}}
\caption{Outage probability for the most protected MODCOD in different environments. Speed: 40 m/s.}
\label{f:outage-40ms}
\end{figure}

Results for the intermediate-tree shadowing, heavy-tree shadowing, open and suburban environments are shown for mobile speeds of $0.1$\,m/s and $40$\,m/s on Figure \ref{f:outage-40ms}. It is easy to conclude that higher speeds dramatically reduce the required SNR for a given outage probability, thus allowing the system to operate with lower SNR values. As an example, note that enforcing an outage probability of $0.1$ in the $0.1$\,m/s case under the heavy-tree shadowing environment requires the received LOS SNR to be larger than $16$\,dB; this value is reduced to $10$\,dB if the speed is increased up to $40$\,m/s.


%

\subsection{Performance of the Proposed Optimized MODCODs}

\begin{figure}
\begin{center}

 \includegraphics[width=.6\textwidth]{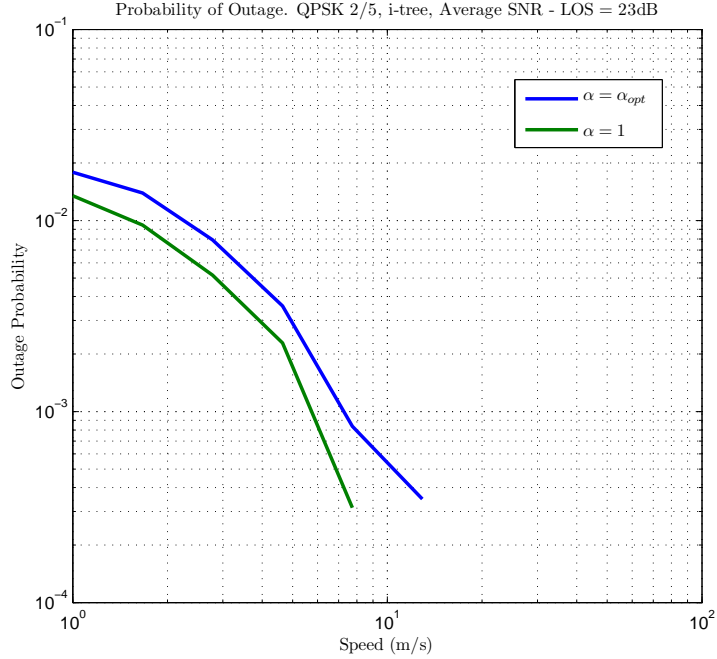}
\caption{Outage probability with the parameters obtained from the optimization (assuming block fading), compared with the SLC case ($\alpha=1$). Target $P_{out} = 0.02$.}

\label{outProbabilityOpt}\end{center}
\end{figure}

We have evaluated the outage and throughput of the forward link with the optimized MODCODs, assuming that the gateway is aware of the statistics of the channel. First of all, we have solved the optimization for a constraint on the outage probability of $P_{out} = 2\cdot10^{-2}$. The LOS SNR was set to a relatively high --and maybe unrealistic-- value, in order to be able to use some higher rate MODCODs. In Figure \ref{outProbabilityOpt}, it can be seen that the outage probability constraint is met for all speeds, and specially for the larger ones. Therefore, the block fading approach turns out to be quite conservative.
 
\begin{figure}
\begin{center}
 \includegraphics[width=.75\textwidth]{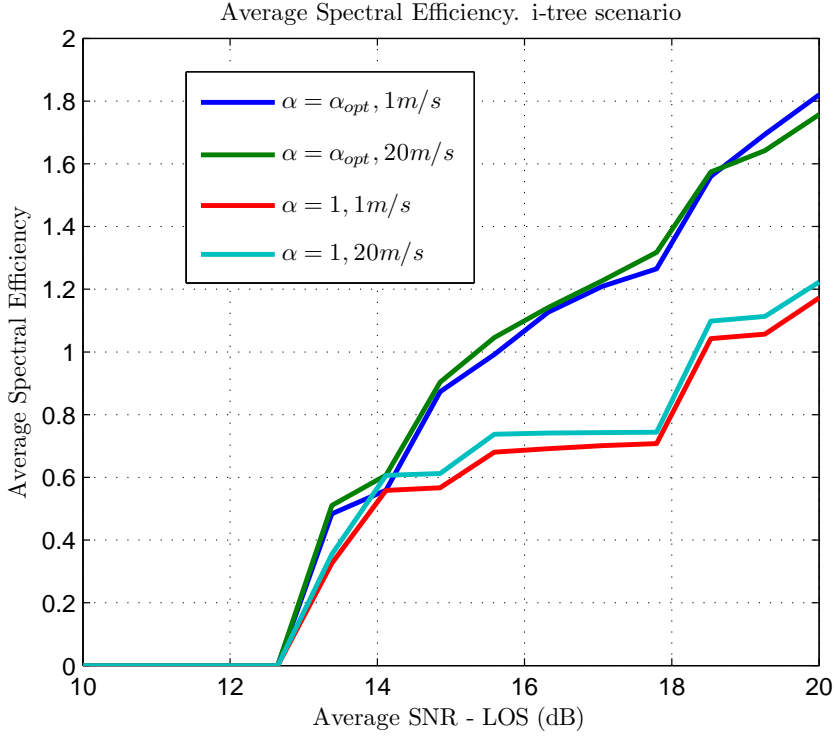}
\caption{Average spectral efficiency with SLC and MLC in ITS environment. Target $P_{out} = 0.1$.}
\label{f:th-i-tree}\end{center}
\end{figure}

Similarly to the results on the outage probability presented in the previous section, we will evaluate the performance of the proposed scheme as a function of the LOS SNR. In this case we have set the outage probability constraint to $0.1$ in order to have a feasible problem for realistic SNR values. In Figure \ref{f:th-i-tree} there is a plot of the ASE as a function of the LOS SNR for the heavy-tree, suburban and intermediate-tree scenarios, respectively, for both MLC and SLC ($\alpha = 1)$. It can be seen that the MLC outperforms SLC for almost every SNR value, while meeting the outage probability constraint due to the selection of the parameter $\alpha$. It is worth remarking that the throughput does not change with the speed in most of the cases, although higher speeds obviously attain a lower outage probability. Note that an ASE equal to zero indicates that the problem is unfeasible for the selected SNR value. The values of $P_{\textnormal{out}}$ obtained stayed below the target value in both cases, with lower values in the case of $v = 20$\,m/s.

\section{Adaptive Coding and Modulation in the Return Link}
\label{sec:rl}
Differently from the forward link, the delay in the return link can be assumed negligible when operating in open-loop mode. For this reason, it is possible to address the design of an ACM scheme based on on-the-fly parameter estimation on the forward link signal. In this section, we will explain how this can be done by making use of the ESM.

\subsection{System Model}
\label{sec:sys_model_rl}
We will assume that the LOS component remains constant within the length of a burst, so that the channel model can be expressed as $\h = \left[h_1 \ h_2 \cdots h_L\right]$ with
\begin{equation}
 \h = \h_{\textnormal{LOS}}+\h_{\textnormal{NLOS}} = \mu\1+\h_{\textnormal{NLOS}},
\end{equation}
so that
\begin{equation}
 h_i\sim\mathcal{C}\mathcal{N}(\mu,2\theta^2)
\end{equation}
where $\mu$ is the mean amplitude in natural units and $2\theta^2$ is the NLOS power. As stated, we will assume that the LOS SNR is known at the transmitter, so that all the uncertainty that we must take into account is on the NLOS component. For this reason, our return link may be simplified to a Rician channel whose time samples are correlated.

Each sample from the SNR vector $\gammab$, which we will denote again $\gamma_i$, will follow a scaled non-central Chi-squared distribution with two degrees of freedom and non-centrality parameter $\lambda$
\begin{equation}\label{eq:distr_gamma}
 \gamma_i\sim\frac{\theta^2}{\sigma^2}\cdot\bar\chi_2^2(\lambda)
\end{equation}
where
\begin{equation}
 \lambda = \left(\frac{\mu}{\theta}\right)^2.
\end{equation}
For simplicity, we will define $\zeta\dot=\theta/\sigma$. As already stated, the MODCOD for each burst will be obtained from its ESM; for the next derivations, we will focus on the approximation of the MIESM by an exponential mapping (EM-ESM).

\subsection{Block Fading}
\label{sec:block_fading_rl}
As in the forward link, when the mobile speed is low, we can approximate $\gamma_1 \approx \gamma_2 \approx \cdots \approx\gamma_L \dot= \gamma$, so that $\gammab = \gamma\1$ and, as a consequence, $\gamma_{\textnormal{eff}} = \gamma$, yielding
\begin{equation}\label{eq:low_speed}
 P\left[\gamma_{\textnormal{eff}}<\gamma_{\textnormal{th}}\right] = F_\gamma(\gamma_{\textnormal{th}}) = 1-Q\left(\frac{\mu}{\zeta},\frac{\sqrt{\gamma_{\textnormal{th}}}}{\zeta}\right)
\end{equation}
where $Q$ is Marcum's $Q$ function and $F_\gamma$ denotes the cumulative distribution function (CDF) of $\gamma$. From (\ref{eq:low_speed}), and given a target outage probability ${\textnormal{out}}$, we would select the highest MODCOD having a threshold below $\gamma_{\textnormal{th}}$, where this value must satisfy
\begin{equation}
 Q\left(\frac{\mu}{\zeta},\frac{\sqrt{\gamma_{\textnormal{th}}}}{\zeta}\right) = 1-P_{\textnormal{out}}.
\end{equation}

\subsection{Beyond Block Fading}
\label{sec:high_speed}
Similarly to the forward link, the block fading approximation will become conservative as the speed increases. The problem is that the pdf of $\gamma_{\textnormal{eff}}$ cannot be obtained in closed form for any speed, not even for large values of $L$, because the random variables $\gamma_i$ are correlated. However, we will show that its pdf can be accurately approximated by that of a log-normal random variable for many speeds of practical interest. Throughout this section, we will show two ways in which this modeling can be exploited.

\subsubsection{Log-normal Approximation}
\label{subsec:lognormal}
\begin{figure}
 \centering
 \includegraphics[width=.65\columnwidth]{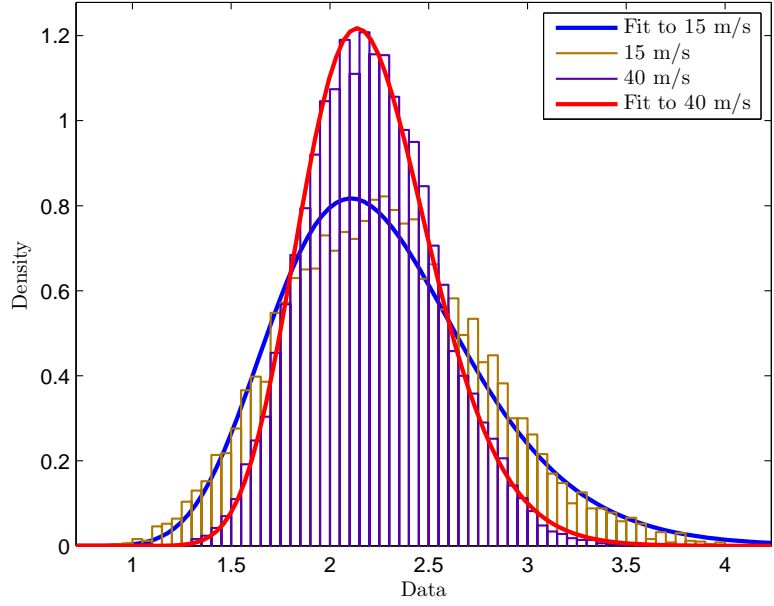}
 \caption{Accuracy of the log-normal approximation. Results have been extracted for ITS at state 2 with $T_{\textnormal{symb}} = 2.7\,\mu$s.}
 \label{fig:lognormal_approximation}
\end{figure}

The mean ESM of a Rician channel is available for the uncorrelated case\cite{rico2012}. Its pdf for a correlated Rician channel lacks of closed form expression, but a number of approximations have been reported in the literature. Here, we will make use of the log-normal approximation proposed in Ref.~\cite{donthi2011}:
\begin{equation}\label{eq:lognormal_approximation}
f_{\gamma_{\textnormal{eff}}}(x) = \frac{1}{x\sqrt{2\pi}\Omega}e^{-\frac{(\log x-M)^2}{2\Omega^2}}
\end{equation}
with
\begin{equation}
	\Omega = \sqrt{\log\left(\frac{m_2}{m_1^2}\right)},
\end{equation}
\begin{equation}
	M = \log m_1-\frac{\Omega^2}{2},
\end{equation}
and
\begin{equation}\label{eq:m1}
	m_1 = \ex\left[\gamma_{\textnormal{eff}}\right]
\end{equation}
and
\begin{equation}\label{eq:m2}
	m_2 = \ex\left[\gamma_{\textnormal{eff}}^2\right].
\end{equation}
The probability of outage would read as
\begin{equation}
 P\left[\gamma_{\textnormal{eff}}<\gamma_{\textnormal{th}}\right] = Q\left(-\frac{\log \gamma_{\textnormal{th}}-M}{\Omega}\right).
\end{equation}
The accuracy of this approximation is shown on Fig.~\ref{fig:lognormal_approximation}, which illustrates a remarkably good fit. In order to make use of this approximation, we would only need to compute, or estimate, the first two moments of $\gamma_{\textnormal{eff}}$, $m_1$ and $m_2$. This could be done, for instance, by using the sample moments obtained from a set of received samples. Nevertheless, we will try to find closed-form expressions for these parameters from those of the distribution of $\h$ (or of $\gammab$). It is important to stress that, for the design of a transmitter, the use of the sample moments should be evaluated anyway, since it would provide a straightforward way of computing the outage probability.

\subsubsection{PDF of the Mutual Information.}
\label{subsec:mi}
In order to obtain some insights on the dependence of $\gamma_{\textnormal{eff}}$ with the distribution of $\h$, we will try to derive closed-form expressions for $m_1$ and $m_2$ in (\ref{eq:m1}) and (\ref{eq:m2}), that is, for the two first moments of the distribution of $\gammaeff$. To accomplish this task, we will focus on characterizing the mutual information $\Phi(\gammaeff)$ and then obtaining the moments of its natural logarithm.

Let us start by characterizing the pdf of the mutual information 
\begin{equation}
 \x\,\dot=\,\Phi(\gammaeff) = \frac{1}{N}\sum_{i=1}^Le^{-\beta\gamma_i}.
\end{equation}
To do so, we will exploit the fact that $L$ will be in general large and resort to a Central Limit Theorem (CLT) for weakly correlated random variables\cite{billingsley95}. Its application, after making some simplifying assumptions on the autocorrelation of the exponentials $\varsigma_i\,\dot=\,e^{-\beta\gamma_i}$ yields
\begin{equation}
 \xi\sim\mathcal{N}(\mu_\xi,\sigma_\xi^2)
\end{equation}
with
\begin{equation}\label{eq:mean_xi}
 \mu_\xi = \ex[\varsigma]
\end{equation}
and 
\begin{equation}\label{eq:var_xi}
 \sigma_\xi^2 = \frac{1}{2f_DT_{\textnormal{symb}}}\ex\left[\left(\varsigma-\ex\left[\varsigma\right]\right)^2\right].
\end{equation}

Before going further, it is worth stressing two points. Firstly, the assumption of almost uncorrelated summands requires the speed to be relatively high; lower speeds will have to be treated as block fading, regardless of how conservative this would be. Secondly, the mutual information will look very different to a Gaussian distribution as the SNR grows higher, because in that case it will tend to one, with no higher values. As a consequence, these derivations, although insightful, may be applicable in a limited range of SNR.

From (\ref{eq:mean_xi}) and (\ref{eq:var_xi}), it is clear that the pdf of the mutual information will be fully characterized if we manage to obtain $\ex[\varsigma]$ and $\ex[\varsigma^2]$. This can be done in the following way: let us recall that
\begin{equation}
 f_\gamma(x) = \frac{1}{2\zeta^2}e^{-\frac{x}{2\zeta^2}-\frac{\lambda}{2}}I_0\left(\frac{\sqrt{\lambda x}}{\zeta}\right),
\end{equation}
then the necessary parameters are obtained from
\begin{equation}\label{eq:mean_varsigma}
 \ex[\varsigma] = \int_0^\infty e^{-\beta x}f_\gamma(x)\, \partial x = \frac{e^{-\frac{\beta\lambda\zeta^2}{1+2\beta\zeta^2}}}{1+2\beta\zeta^2},
\end{equation}
\begin{equation}
  \ex[\varsigma^2] = \int_0^\infty \left(e^{-\beta x}\right)^2 f_\gamma(x)\, \partial x = \frac{e^{-\frac{2\beta\lambda\zeta^2}{1+4\beta\zeta^2}}}{1+4\beta\zeta^2}.
\end{equation}
Note that substitution in (\ref{eq:var_xi}) yields
\begin{equation}
 \sigma_\xi^2 = \frac{\frac{e^{-\frac{2\beta\lambda\zeta^2}{1+4\beta\zeta^2}}}{1+4\beta\zeta^2}-\frac{e^{-\frac{2\beta\lambda\zeta^2}{1+2\beta\zeta^2}}}{(1+2\beta\zeta^2)^2}}{2f_DT_{\textnormal{symb}}},
\end{equation}
so that finally
\begin{equation}
 \Phi(\gammaeff)\sim\mathcal{N}\left(\frac{e^{-\frac{\beta\lambda\zeta^2}{1+2\beta\zeta^2}}}{1+2\beta\zeta^2},\frac{\frac{e^{-\frac{2\beta\lambda\zeta^2}{1+4\beta\zeta^2}}}{1+4\beta\zeta^2}-\frac{e^{-\frac{2\beta\lambda\zeta^2}{1+2\beta\zeta^2}}}{(1+2\beta\zeta^2)^2}}{2f_DT_{\textnormal{symb}}}\right).
\end{equation}

After the full characterization of the pdf of the mutual information, it is still required to obtain the moments of the transformation $g(\xi) = -\frac{1}{\beta}\log\xi$ from the pdf of $\xi$. They can be computed as

\begin{equation}
 m_1 = \frac{1}{\sqrt{2\pi}\beta\sigma_\xi}\int_0^\infty\log(x)e^{-\frac{(x-\mu_\xi)^2}{2\sigma_\xi^2}}\,\partial x
\end{equation}
and
\begin{equation}
m_2 = \frac{1}{\sqrt{2\pi}\beta\sigma_\xi}\int_0^\infty\log(x)^2e^{-\frac{(x-\mu_\xi)^2}{2\sigma_\xi^2}}\,\partial x.
\end{equation}
Although the above integrals lack of closed-form solution, they are simple to evaluate numerically.

At this point, we already have analytical expressions to compute the distribution of the ESM (by means of the log-normal approximation with the above derived moments) and, as a consequence, to design an ACM loop that tries to guarantee a given outage probability; this can be done by simply using the CDF of the ESM, and possibly adding some security margin to take into account the imperfections in the approximation. It is worth remarking that the obtained closed-form expressions for the moments hold for high speed only; however, a transmitter might decide not to use them and estimate empirically the sample ESM moments instead.

\subsection{Simulation Results for the Return Link}
In this section we report the results obtained with the two previously reported approximations: block fading for low speed and log-normal distribution for higher speeds. Results have been extracted with a symbol period of $2.7\,\mu$s for an ITS area, fixing a target outage probability of $0.01$; the MODCOD set used can be found in Table \ref{tbl:modcod}, and the frequency has been set to $f = 2.2$\,GHz. Bursts contain $8100$ symbols each, and the last $123$ bursts are used at each time instant to estimate the moments of the ESM from estimations of the SNR; the LOS component is kept fixed within each state

\begin{figure}
\hspace{-.5cm}
 \includegraphics[width=.56\columnwidth]{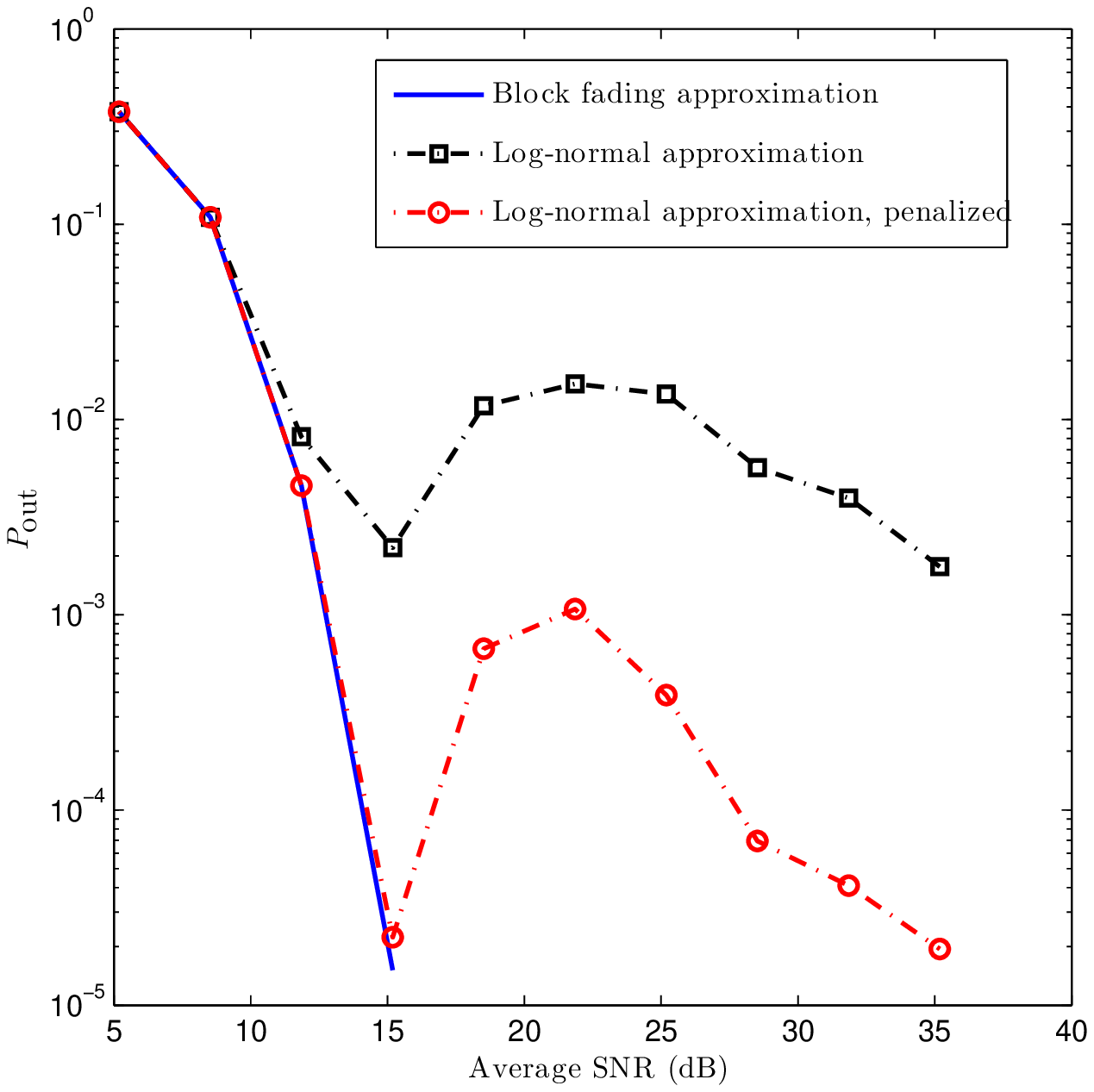}\hspace*{-.9cm}
\includegraphics[width=.56\columnwidth]{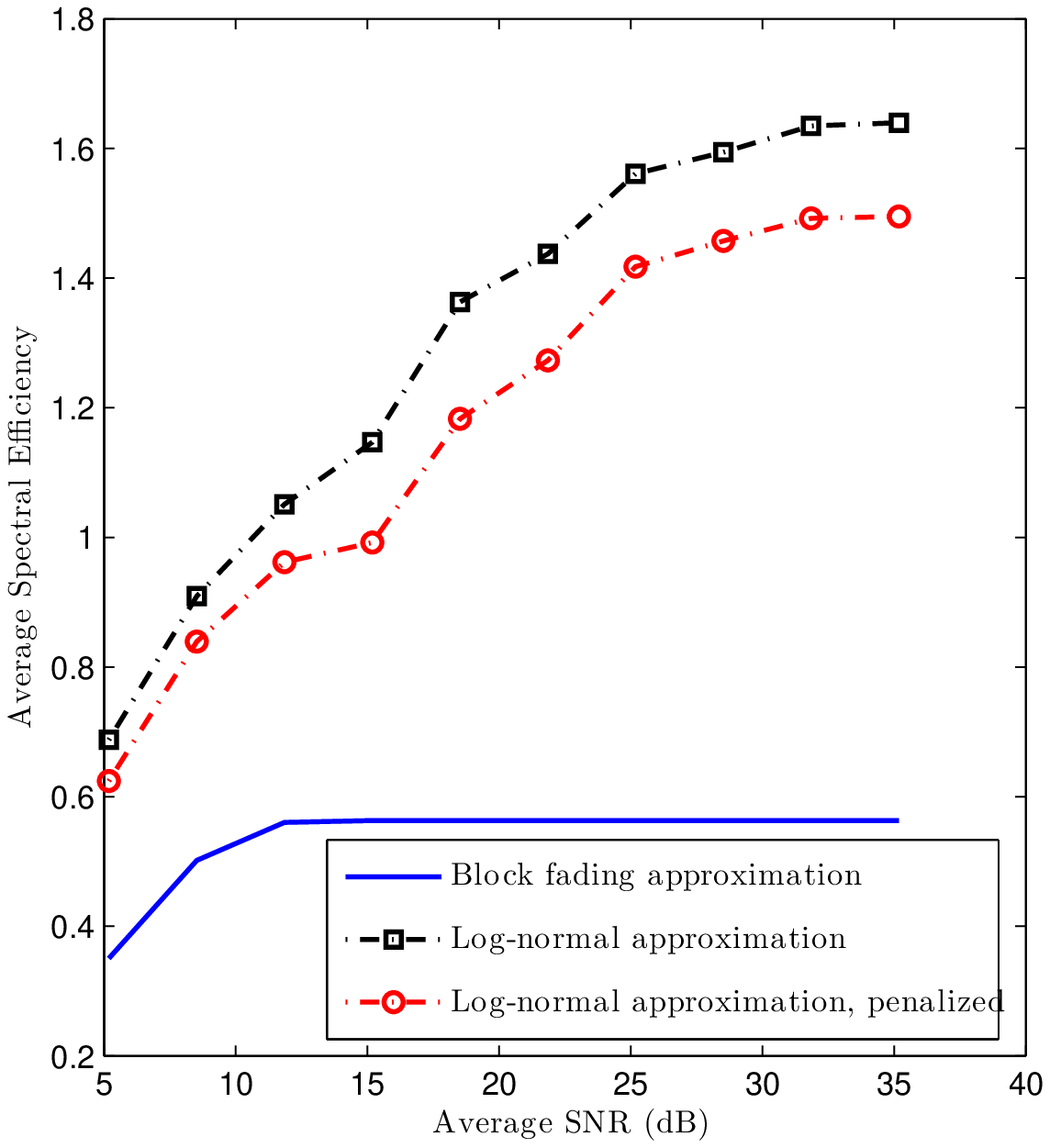}
 \caption{Evolution of the probability of outage and the spectral efficiency with respect to the average SNR. Target $P_\textnormal{out}=0.01$ and $2\cdot10^6$ codewords per point. ITS area, state 2.}
 \label{fig:pout_snr_rl1}
\end{figure}

The obtained results can be seen on Fig.~\ref{fig:pout_snr_rl1}, where the improvement in terms of spectral efficiency when using the log-normal approximation is clear. Note that because of using past samples to estimate the moments, there might be small impairments in the target outage probability; if these are too high, then a one-MODCOD penalization might be used as illustrated in the figure. The obtained results still show an increase in the average spectral efficiency, although it is worth noticing that better performances would be achieved if a more granular MODCOD set was available.

\section{Conclusions}\label{sec:conclusions}
Adapting the transmission rate in an LMS channel is a challenging task because of the relatively fast time variations, of the long delays involved, and of the difficulty in mapping the parameters of a time-varying channel into communication performance. In this paper, we have proposed two strategies for dealing with this impairments, namely, multi-layer coding (MLC) in the forward link and log-normal modeling --based on open-loop estimations-- in the return link. Both strategies rely on an effective SNR mapping, borrowed from the physical-layer abstraction literature, as a tool for predicting the link performance. We have shown that, in both cases, it is possible to increase the average spectral efficiency while at the same time keeping the outage probability under a given threshold.

\section*{Acknowledgments}\label{sec:ack}

The authors are very grateful to Riccardo de Gaudenzi, from ESA, for his valuable comments and suggestions. 
Work supported by the European Space Agency within the SatNEx III Network of Experts  under ESTEC Contract No. 23089/10/NL/CLP.


\end{document}